\documentclass[11pt,a4paper]{article}
\usepackage{jheppub}
\usepackage{amsfonts,amsmath,hyperref,url, color}
\usepackage{eurosym}
\usepackage{braket}
\usepackage{graphicx}

\usepackage{xcolor}

\def\cC{{\cal C}}

\def\cW{{\cal W}}
\def\cT{{\cal T}}

\title{%
  BMS algebras in 4 and 3 dimensions, their quantum deformations and duals}

\author[a]{Andrzej Borowiec}
\author[a]{Lennart Brocki}
\author[a,b]{Jerzy Kowalski-Glikman}
\author[a]{Josua Unger}
\affiliation[a]{Institute for Theoretical Physics, University of Wroc\l{}aw, pl.\ M.\ Borna 9, 50-204 Wroc\l{}aw, Poland}
\affiliation[b]{National Centre for Nuclear Research, Pasteura 7, 02-093 Warsaw, Poland}
\emailAdd{andrzej.borowiec@uwr.edu.pl, lennart.brocki@uwr.edu.pl,\\
	 jerzy.kowalski-glikman@uwr.edu.pl, unger.josua@uwr.edu.pl}

\abstract{BMS symmetry is a symmetry of asymptotically flat spacetimes in vicinity of the null boundary of spacetime and it is expected to play a fundamental role in physics. It is interesting therefore to investigate the structures and properties of quantum deformations of these symmetries, which are expected to shed some light on symmetries of quantum spacetime. In this paper we discuss the structure of the algebra of extended BMS symmetries in 3 and 4 spacetime dimensions, realizing that these algebras contain an infinite number of distinct Poincar\'e subalgebras, a fact that has previously been noted in the 3 dimensional case only. Then we use these subalgebras to construct an infinite number of different Hopf algebras being quantum deformations of the BMS algebras. We also discuss different types of twist-deformations and the dual Hopf algebras, which could be interpreted as noncommutative, extended quantum spacetimes.}
\keywords{BMS symmetry, quantum deformations of spacetime symmetries}
\begin{document}
\maketitle

\section{Introduction}

In the recent years we witness a revival of interest and a remarkable progress in understanding the structure of the gravitational field in the vicinity of the future (past) null boundary of asymptotically flat spacetime. The original, surprising result of Bondi, van der Burg, Metzner, and Sachs \cite{Bondi:1962px}, \cite{Sachs:1962wk}, \cite{Sachs:1962zza} that the symmetry of gravity near the null boundary is much larger than the expected Poincar\'e group and in fact is an infinite dimensional BMS group, whose algebra we will denote $\mathcal B_{\text{BMS}}^4$, consisting of Lorentz transformations and infinitely many commuting supertranslations was generalized and understood from many different perspectives. As it was shown recently the BMS group is also a symmetry of spacelike infinity \cite{Henneaux:2018cst} if some special parity conditions are adopted, effectively superseding Poincar\'e symmetry and making BMS symmetry the fundamental symmetry of asymptotically flat spacetime. It was also argued \cite{Barnich:2010eb} that the original BMS symmetry could be generalized so as to include an infinite number of superrotations and additional supertranslations. It was also recently realized (see \cite{Strominger:2017zoo}, \cite{Ashtekar:2018lor} for comprehensive reviews) that there exist deep interrelations between the BMS group and the infrared behavior of QFT described by Weinberg's soft theorems \cite{Weinberg:1965nx}, which can be understood as Ward identities for supertranslations \cite{Strominger:2013jfa}, \cite{He:2014laa}. As it also turned out, the gravitational memory effect \cite{Christodoulou:1991cr} is ultimately related to the two \cite{Strominger:2014pwa}.
Furthermore, it has recently been claimed that the BMS group should be also relevant in the description of the neighbourhood of the black hole horizon and as a consequence should help in the understanding of the black hole information paradox \cite{Hawking:2016msc}, \cite{Hawking:2016sgy}.

In the recent paper \cite{Borowiec:2018rbr} we constructed the $\kappa$-deformation of the BMS algebra (quantum deformations of BMS algebra were also considered in \cite{Alessio:2019cch}) $\mathcal B^4$ proposed by Barnich and Troessaert \cite{Barnich:2010eb}, which in addition to the infinitely many supertranslations contains an infinite number of superrotations, generalizing the Lorentz sector of the original BMS algebra
$\mathcal B_{\text{BMS}}^4$. The interest in such a deformation is at least threefold.

First, it is believed that quantum deformations of spacetime symmetries could shed some light on the elusive quantum gravity theory. This especially applies to the case of $\kappa$-deformation  \cite{Lukierski:1991pn}--\cite{Majid:1994cy}, in which the deformation parameter has the physical dimension of mass, and therefore can be naturally identified with the Planck mass. It is well established that the deformation of space-time symmetries with a deformation parameter which has the dimension of mass is a direct consequence of (quantum) gravity in 2+1 spacetime dimensions \cite{Freidel:2005me}, \cite{Cianfrani:2016ogm}, \cite{Dupuis:2020ndx} and is expected to take place in 3+1 dimensions as well \cite{AmelinoCamelia:2011bm}.  These investigations concern the Poincar\'e symmetry, but most likely could (and should) be extended to the case of the BMS symmetry. It can be claimed that quantum gravity in the flat spacetime limit could exhibit itself under the guise of a deformation of BMS symmetry. If this is the case, as argued in \cite{Borowiec:2018rbr}, the quantum deformation could provide the required way around the no go arguments put forward in the series of papers \cite{Mirbabayi:2016axw}--\cite{Javadinazhed:2018mle}, which indicate that the original scenario of Hawking, Pope, and Strominger of solving the black hole information paradox fails as a result of the fact that the soft and hard modes of Hawking radiation completely decouple and that the former do not carry any physical information. Contrary to that in the deformed case these modes cannot decouple as a consequence of the presence of a non-trivial coproduct, which makes the structure of multiparticle states much more complicated than in the standard, undeformed  theory.

Second, $\kappa$-deformations yield predictions for various phenomenological effects that are in principle observable. This includes deformed dispersion relations or altered Leibnitz rules for the addition of momenta. On this ground one can not only distinguish the deformed and undeformed case but potentially also between various different versions of deformations. As we will show, some of these versions can be excluded in the BMS setting from purely mathematical reasoning, e.g. the well-known time-like $\kappa$-Poincar\'e. Others, like the lightcone $\kappa$-Poincar\'e, its Jordanian part and the deformation from an abelian twist can be constructed and compared.

Third, the investigation of deformations of infinite-dimensional Lie algebras is an interesting subject on its own. Their mathematical structure and formalism are less developed than in the standard finite-dimensional case. This can be observed already on the level of the Witt algebra \cite{HuWang07}, which is an important constituent of BMS3 and BMS4 algebras connected with superrotations. Fortunately enough, some structures and constructions generalize directly from the finite-dimensional case.  As an example, one can consider triangular Lie bialgebras (provided by solutions of the classical Yang-Baxter equation) and the corresponding quantized enveloping algebras obtained by Drinfeld’s two-cocycle twist techniques \cite{Drinfeld:1983rx,Drinfeld2}. But there are some remarkable differences. Triangular classical $r$-matrices belong to some finite-dimensional subalgebras which might be isomorphic. However, the corresponding infinite-dimensional Lie bialgebras are not. This yields an abundance of bialgebra structures and their quantized enveloping algebras (quantum groups) in the infinite-dimensional case. Also, the problem of classifying such bialgebra structures is much more difficult and, apart from the half Witt algebra, it is not complete. Even more striking differences appear when dualizing the bialgebra structure. For the infinite-dimensional case, this requires developing entirely new tools in the form of sequences satisfying linearly recursive relations (cf. Appendix B and references therein).

The aim of the present paper is to further investigate the $\kappa$-deformation of BMS algebra, both in 3+1 and 2+1 spacetime dimensions. Our interest here will be particularly focused on the dual of the BMS algebra, which could be thought of as an algebraic description of space-time. Indeed, it is well known that in the case of the Poincar\'e algebra the (algebraic) dual of its translational part is the Minkowski space, which has a natural interpretation of being the flat spacetime. In the case of $\kappa$-deformed Poincar\'e algebra, the dual to its translational sector is the non-commutative $\kappa$-Minkowski space \cite{Lukierski:1993wx}, \cite{Majid:1994cy}. Not only does this result fit into the general expectation that quantum gravity makes spacetime non-commutative (even in the flat space limit), but it is also a demonstration of the general fact that the nontrivial coproduct in one algebraic sector (in momentum space in this case) goes hand in hand with non-commutativity of the dual one. It is therefore of interest to explore what will be the generalization of this result to the case of the BMS symmetry.

The plan of the paper is as follows. In section \ref{sec2} we recall some basic facts about BMS symmetry and its algebra. We then notice that the BMS algebra $\mathcal B^4$ in 3+1 dimensions contains an infinite number of subalgebras isomorphic to Poincar\'e algebra. This, as far as we can say, previously unnoticed fact is presumably related to the presence of an infinite number of inequivalent, flat spacetime vacua in asymptotically flat gravity. From our perspective it is of crucial relevance, because our $\kappa$-deformation of BMS algebra uses its Poincar\'e subalgebra as the starting point. %\textcolor{red}{
	An analogous, but not completely identical, structure has been noted in the case of the 2+1 dimensional BMS algebra $\mathcal B^3$ in \cite{Barnich:2014kra}. We furthermore demonstrate in 2+1 dimensions that each of these subalgebras is the isometry of a distinct solution of the vacuum Einstein equations, which turns out to be a spacetime with a conical singularity, %\textcolor{red}{ 
	a conclusion that has been reached previously in \cite{Barnich:2015uva} in a slightly different language. % }
The presence of infinitely many such algebras results in infinitely many distinct, albeit similar,  deformations of the  BMS algebra generalizing the result of \cite{Borowiec:2018rbr}. In section \ref{sec3} we describe the deformations of the 3 dimensional algebra $\mathcal B_{\text{ext}}^3$ and its dual. After examining a different example of deformation in 3 dimensions given by an abelian twist in section \ref{sec5}, in section \ref{sec6} we describe the deformations of the 4 dimensional BMS algebra and its dual. Section \ref{sec7} is devoted to a short summary and conclusions. Finally, three appendices describe some more technical mathematical results.

\section{Symmetries of asymptotically flat spacetimes}\label{sec2}

In this section we will present the BMS algebra and discuss its structure. We start with the 4-dimensional case, which is more familiar, and in the next subsection we turn to the simpler 3-dimensional case.

\subsection{4d gravity}

As was shown in the work of Bondi, van der Burg, Metzner and Sachs (BMS) \cite{Bondi:1962px} \cite{Sachs:1962wk} \cite{Sachs:1962zza} the symmetry transformations that leave the form of an
asymptotically flat metric invariant do not just include translations, rotations and boosts but also so-called supertranslations and, following an extension proposed by Barnich and Troessaert \cite{Barnich:2010eb}, also superrotations. Here we define asymptotic flatness using coordinates in the Bondi gauge
\begin{align}
	g^{uu}=0,\quad g^{uA}=0,\quad \partial_r\text{det}(g_{AB}/r^2)=0.
\end{align}
The first condition implies that the normal vector of hypersurfaces defined by $u=const.$, $n^{\mu}=g^{\mu\nu}\partial_{\nu}u$, is null and thus $u$ is labeling null hypersurfaces. Angular coordinates $x^A$ are defined such that the directional derivative along $n^{\mu}$ vanishes, $n^{\mu}\partial_{\mu}x^A=0$, and $r$ is defined to be the luminosity distance. After lowering the indices these conditions are $g_{rr}=g_{rA}=0$. We define asymptotic flatness near null infinity by introducing boundary conditions such that in the limit $r\rightarrow\infty,\; u=const.$ one obtains the Minkowski metric. Note, however, that in the literature different notions of asymptotic flatness exist, see \cite{Ruzziconi:2020cjt} for a review. These asymptotic boundary conditions should not be too restrictive, however, to still include all physical spacetimes such as black hole mergers with gravitational radiation. Following the analysis of BMS, boundary conditions are defined as
\begin{equation}
	g_{u u}=-1+\mathcal{O}\left(r^{-1}\right), \quad g_{u r}=-1+\mathcal{O}\left(r^{-2}\right), \quad g_{u A}=\mathcal{O}\left(r^{0}\right), \quad g_{A B}=r^{2} \gamma_{A B}+\mathcal{O}(r)
\end{equation}
and the class of allowed metrics then reads \cite{Bondi:1962px}--\cite{Sachs:1962zza}
\begin{align}\label{bondigauge}
ds^2 =& -du^2 -2 du dr + 2r^2 \gamma_{z\bar z} dz d \bar z\nonumber \\
&  + \frac{2 m_B}{r} du^2  + r C_{zz} dz^2 + r C_{\bar z\bar z} d\bar z^2 + D^z C_{zz} du dz +  D^{\bar z} C_{\bar z \bar z} du d\bar z \nonumber \\
& + \frac{1}{r} \left( \frac{4}{3} (N_z  + u \partial_z m_B) - \frac{1}{4} \partial_z (C_{zz} C^{zz}) \right) du dz + c.c. + \text{subleading}
\end{align}
with $u=t-r$ being the retarded time and $z=e^{i\phi}\cot(\theta/2),\,\bar z=z^{\ast}$ (complex) stereographic coordinates of the sphere,  $\gamma_{z\bar z}=\frac{2}{(1+z\bar z)^2}$ the unit round metric on the 2-sphere and $D_A$ the covariant derivative with respect to $\gamma_{z\bar z}$.
Here, $m_B$ is the Bondi mass aspect, $N_A$ the angular momentum aspect and $C_{AB}$ is traceless, $\gamma^{AB}C_{AB}=0$, and symmetric. Thus it has two polarization modes and contains the information about gravitational radiation near $\mathcal I^+$. All three functions are defined in the asymptotic region and have no $r$-dependence.

The residual gauge transformations preserving the Bondi gauge are infinitesimal diffeomorphisms $\xi$ obeying
\begin{equation}
	\mathcal{L}_{{\xi}} g_{r r}=0, \quad \mathcal{L}_{{\xi}} g_{r A}=0, \quad \mathcal{L}_{{\xi}} \partial_{r} \operatorname{det}\left(g_{A B} / r^{2}\right)=0
\end{equation}
and it is further required that  the asymptotic behavior is preserved leading to the conditions
\begin{equation}\label{falloff}
	\mathcal{L}_{{\xi}} g_{u u}=\mathcal{O}\left(r^{-1}\right), \quad \mathcal{L}_{{\xi}} g_{u r}=\mathcal{O}\left(r^{-2}\right), \quad \mathcal{L}_{{\xi}} g_{u A}=\mathcal{O}\left(r^{0}\right), \quad \mathcal{L}_{{\xi}} g_{A B}=\mathcal{O}(r).
\end{equation}
Solving these requirements one obtains the BMS generators in terms of two functions on the sphere $f(z, \bar z)$ and $R^A(z, \bar z)$
\begin{equation}\label{bmsgenerator}
	\begin{aligned}
		\xi(f, R)=&\left[f+\frac{u}{2} D_{A} R^{A}+o\left(r^{0}\right)\right] \partial_{u} \\
		&+\left[R^{A}-\frac{1}{r} D^{A} f+o\left(r^{-1}\right)\right] \partial_{A} \\
		&+\left[-\frac{r+u}{2} D_{A} R^{A}+\frac{1}{2} D_{A} D^{A} f+o\left(r^{0}\right)\right] \partial_{r}
	\end{aligned}
\end{equation}
which can be understood as "asymptotic Killing vectors" since they leave the metric only asymptotically invariant. While $f$ is unconstrained the last condition in \eqref{falloff} imposes that $R^A$ has to obey the conformal Killing equation on the 2-sphere
\begin{equation}
	D_{A} R_{B}+D_{B} R_{A}=\gamma_{A B} D_{C} R^{C}.
\end{equation}
This equation imposes the conditions $\partial_z R^{\bar z}=0$ and $\partial_{\bar z}R^{z}=0$, so that  $R^{z}(z)$ is an holomorphic function and $R^{\bar z}(\bar z)$ antiholomorphic.

In leading order these Killing vectors form the 4-dimensional BMS algebra $\mathcal B^4$, which is an infinite-dimensional algebra with the bracket
\begin{equation}\label{bms4}
	\xi(\hat{f}, \hat{R})=\left[\xi(f, R),\xi(f^{\prime}, R^{\prime})\right] \Longrightarrow\left\{\begin{array}{l}
		\hat{f}=R^{A} D_{A} f^{\prime}+\frac{1}{2} f D_{A} R^{\prime A}-R^{\prime A} D_{A} f-\frac{1}{2} f^{\prime} D_{A} R^{A} \\
		\hat{R}^{A}=R^{B} D_{B} R^{\prime A}-R^{\prime B} D_{B} R^{A}
	\end{array}\right..
\end{equation}
As is explained in \cite{Barnich:2010eb} in order for the vector fields \eqref{bmsgenerator} to be a faithful representation of $\mathcal B^4$ in all orders, one needs to modify the standard Lie bracket of the vector fields, so as to take into account the change of the metric components the vector fields \eqref{bmsgenerator} implicitly depend on. It can be seen from \eqref{bms4} that $T=\xi(f,0)$ form an abelian ideal of the algebra. These generators are referred to as supertranslations since they generalize the Poincar\'e translations.
The generators $l=\xi(0,R)$ contain the Lorentz algebra and possible extensions of it and will be referred to as superrotations.

It is interesting to observe what happens when one acts with a supertranslation on the Minkowski vacuum, which is characterised by $m_B=C_{zz}=N_{zz}=0$, and in which case one finds \cite{Strominger:2017zoo}
\begin{align}
	\mathcal L_f m_B=0,\quad \mathcal L_f N_{zz}=0,\quad \mathcal L_f C_{zz}=-2D_z^2f.
\end{align}
The supertranslated vacuum still has zero Bondi mass and Bondi News but it obtains a finite $C_{zz}$ term. Furthermore, for the curvature to vanish we need to have \cite{Strominger:2017zoo}
\begin{align}
	C_{zz}=-2D_z^2C(z,\bar z),
\end{align}
where the function $C$ transforms as
\begin{align}
	\mathcal L_f C=f.
\end{align}
The fact that the supertranslations do not leave the vacuum invariant, even though they are an asymptotic symmetry, can be viewed as the spontaneous breaking of that symmetry and $C$ is then the Goldstone boson labeling inequivalent gravitational vacua, see also \cite{Compere:2016jwb}. We will return to this important point later in this section where we observe an analogous effect on the level of Poincar\'e subalgebras of $\mathcal B^4$.

For our further discussion it will be useful to expand supertranslations and superrotations in the basis of $z$, $\bar z$ monomials
\begin{align}
f_{mn}=\, \frac{z^m \bar z^n}{1 + z \bar z},\quad R^z_{n}= -\, z^{n+1},\quad R^{\bar z}_{n} = -\bar z^{n+1}.
\end{align}
In terms of the basis vectors $T_{mn}=\xi(f_{mn},0),\,l_n=\xi(0,R_{n}^z),\, \bar l_n=\xi(0,R_{n}^{\bar z})$ the algebra \eqref{bms4} takes the form
\begin{align}\label{bms}
[l_m, l_n] = (m-n)l_{m+n}, \quad [\bar l_m, \bar l_n ] = (m-n) \bar l_{m+n}, \quad [l_m, \bar l_n]= 0  \\
[l_l, T_{m,n}] = \left(\frac{l+1}{2} -m\right)T_{m+l, n}, \quad [\bar l_l, T_{m,n}] = \left(\frac{l+1}{2} -n\right)T_{m, n+l}.
\end{align}
One can identify the Poincar\'e translational generators in Cartesian coordinates as
\begin{align}
	P_0+P_3& = T_{11}, \quad P_0-P_3 = T_{00},\nonumber \\
	P_1 & = T_{10} + T_{01}, \quad P_2 = i(T_{10} - T_{01}),\label{Ptrans}
\end{align}
and the boosts and rotations as
\begin{align}
	&J_1 = -\frac 12(l_1+l_{-1}+\bar l_1+\bar l_{-1}),\quad K_1=\frac{i}{2}(l_1+l_{-1}-\bar l_1-\bar l_{-1})\nonumber \\
	&J_2 = \frac i2(l_1-l_{-1}+\bar l_1+\bar l_{-1}),\quad K_2=\frac 12(l_1-l_{-1}+\bar l_1-\bar l_{-1})\nonumber\\
	&J_3 = l_0+\bar l_0,\quad K_3 = l_0-\bar l_0.\label{Prot}
\end{align}
Historically, the BMS algebra $\mathcal B_{\text{BMS}}^4$ of Bondi, van der Burg, Metzner, and Sachs was first defined as the semi-direct sum of the infinite-dimensional abelian algebra of supertranslations $\mathbf s$ and the Lorentz algebra $\mathbf{so}(3,1)$, i.e. $l_n,\;n=0,\pm 1$, $T_{m,n}\; n \in \mathbb{N}$
\begin{align}
	\mathcal B_{\text{BMS}}^4=\mathbf{so}(3,1)\oplus_S \mathbf s.
\end{align}
Two extensions of the $\mathcal B_{\text{BMS}}^4$ algebra have been proposed in recent years. In \cite{Barnich:2009se} the algebra is extended by including all $l_n,\; n \in \mathbb{Z}$, which implies the inclusion of $m,n\in \mathbb{Z}$ the labels of $T_{m,n}$ as well. The generators of the extended Lorentz algebra $l_n$ are referred to as superrotations. The resulting algebra is referred to as extended BMS algebra and is the semi-direct sum of two copies of infinitesimal diffeomorphisms on $S^1$ and extended supertranslations $\mathbf{s}^\ast$
\begin{align}
	\mathcal B_{\text{ext}}^4 = \left(\mathbf{\text{Diff}}(S^1)\oplus \mathbf{\text{Diff}}(S^1)\right)\oplus_S\mathbf{s^{\ast}}.
\end{align}
It is this extension we will concern ourselves with throughout this article as $\mathcal B^4$.

In \cite{Campiglia:2014yka} a different extension has been proposed in which $R^A$ is not constrained by the conformal Killing equation. %\textcolor{red}{
To this end the condition \eqref{falloff} is softened to
\begin{align}
	\mathcal L_{\xi} g_{AB}=O(r^2)
\end{align}
which allows the leading order of $g_{AB}$ to fluctuate such that it is not identical to the round metric anywhere. In particular this means that in the large $r$ limit one does not obtain the Minkowski metric and one is thus working with a different notion of asymptotic flatness, which is explained in detail in \cite{Campiglia:2014yka}. Since $R^A$ is not constrained the resulting algebra is
\begin{align}\label{genB}
	\mathcal B_{\text{gen}}^4=\text{Diff}(S^2)\oplus_S \mathbf s
\end{align}
and is referred to as generalized BMS algebra.

Contrary to the classical BMS algebra $\mathcal B_{\text{BMS}}^4$ the extended BMS algebra \eqref{bms} contains an infinite number of (overlapping) subalgebras generated by the sets of 10 generators
$$
\left\{ l_0, l_{1-2m},\bar l_0, \bar l_{1-2m}, T_{m, m}, T_{1-m, m}, T_{m, 1-m}, T_{1-m, 1-m}\right\}
$$
with $m \in \mathbb{Z}$, which are isomorphic to the Poincar\'e algebra after rescalling
\begin{align}
l_i \rightarrow \frac{l_i}{n}, \quad \bar l_i \rightarrow \frac{\bar l_i}{n},
\end{align}
where $n = 1-2m$.
Note, however, that this is not an automorphism on the entire $\mathcal{B}^4_{\text{ext}}$, e.g.
\begin{align}
[l_p',l_q']=\frac{1}{n}(p-q)l'_{p+q},\quad l_p'=\frac{1}{n}l_p,
\end{align}
which coincides with \eqref{bms} only for $n=1$.

It is tempting to associate these Poincar\'e subalgebras with algebras of symmetries of inequivalent vacua in 4d asymptotically flat spacetime \cite{Compere:2016jwb}, \cite{Ashtekar:2018lor}. It is important to note, however that the Poincar\'e subalgebras  constructed above necessarily contain superrotations. On the other hand,  in  the asymptotically flat gravity (or from the field theoretical soft gravitons perspective) the vacua are found using only the supertranslations \cite{Compere:2016jwb}, and superrotations seemingly do not play any role. We will address these issues in details in a separate publication; below we briefly discuss this issue in the case of 3 dimensions.

\subsection{3d gravity}

In the case of 3d gravity the discussion of asymptotically flat spacetimes is similar to the 4d case. Analogously to the 4d case asymptotically flat spacetimes are defined to have the following form at large $r$
\begin{align}\label{3dfalloff}
ds^2 = \left(-1+\frac{\bar g_{uu}}{r}\right)du^2+2\left(-1+\frac{\bar g_{ur}}{r^2}\right)dudr+\left(h(u,z)+\frac{\bar g_{uz}}{r}\right)dudz\nonumber\\-\left(\frac{r^2}{z^2}+\bar{g}_{z z} r\right) dz^2 + ...,
\end{align}
where the barred metric components are functions of $u=t-r$ and $z=e^{i\phi}$. The asymptotic Killing vectors for supertranslations can be parametrized by $z^n$ and read \cite{Oblak:2016eij}
\begin{align}\label{3dsupertran}
T_n=z^n\partial_u-nz^{n+1}\int_r^{\infty}\frac{dr'}{r'^2}g_{r'u}\partial_z+\left(rnz^n\int_r^{\infty}\frac{dr'}{r'^2}(ng_{r'u}+\partial_z g_{r'u})+\frac{nz^{n+1}}{r}g_{zu}\right)\partial_r,
\end{align}
where the unbarred metric components are the full, $r$-dependent components, and using their large $r$ behavior according to \eqref{3dfalloff} the supertranslation generators can be written as
\begin{align}
T_n=z^n\partial_u+\left(\frac{nz^{n+1}}{r}+O(1/r^2)\right)\partial_z-\left(n^2z^n+O(1/r)\right)\partial_r.
\end{align}
For the superrotations the Killing vectors are, using the same parametrization,
\begin{align}\label{3dsuperrot}
l_{n}= & i\bigg(n u z^{n} \partial_{u}-\left(n r z^{n}-uzn^{2}\partial_{z}\left[z^{n} \int \frac{d r^{\prime}}{r^{\prime 2}} g_{r^{\prime} u}\right]+\frac{n^2z^{n+1}}{r}g_{uz}\right)\partial_r \nonumber \\
&+z^{n+1}\left(1-u n^{2}\int \frac{d r^{\prime}}{r^{\prime 2}} g_{r^{\prime} u}\right) \partial_{z}\bigg)
\end{align}
and using the expansion at large $r$
\begin{equation}
l_n = iz^n\left(nu\partial_u-\left(rn+O(r^0)\right)\partial_r+\left(z+O(r^{-1})\right)\partial_z\right).
\end{equation}
Setting $n=0,\pm 1$ one obtains the generators of the three standard Poincar\'e translations, one rotation and two boosts which are related as
\begin{align}
&\partial_t = T_0,\quad \partial_x = T_1+T_{-1},\quad \partial_y=i(T_1-T_{-1}) \\
&\partial_{\phi}=l_0,\quad x\partial_t-t\partial_x=l_1+l_{-1},\quad y\partial_t-t\partial_y=i(l_1-l_{-1}).
\end{align}
At leading order the Killing vectors, under the standard Lie bracket,
form a representation of the abstract infinite-dimensional BMS3 algebra  with (real: $X^\dagger=-X$) anti-Hermitian generators (see Appendix \ref{AppendixA})
\begin{align}\label{bms3d}
[l_m, l_n ] = (m-n) l_{m+n}, \quad [l_m, T_n] = (m-n) T_{m+n}\,,
\quad m, n\in\mathbb{Z}\,,
\end{align}
which we refer to as $\mathcal B^3$.
In order for the Killing vectors to form a faithful representations of the BMS algebra at all orders the Lie bracket has to be modified \cite{Barnich:2010eb}. This modification arises because the $T_m$ and $l_m$ leave the metric only asymptotically invariant and since they themselves depend on the metric components one has to take into account the change of the Killing vectors induced by the change in the metric.
\newline
 Comparing with the 3d Poincar\'e algebra in lightcone coordinates
 \footnote{Following physicist convention we are using Hermitian generators.}
\begin{align}\label{pc1}
[ M_{+ 1}, M_{-1} ] & = i\eta_{1 1} M_{+-}\,, \quad [M_{+-}, M_{\pm 1}] = \mp i \eta_{+ -}M_{\pm 1} \,, \\
[M_{+ -}, P_{\pm} ] & = \mp i \eta_{+ -} P_{\pm}\,, \quad [M_{\pm 1}, P_1] = -i \eta_{1 1} P_{\pm}\,, \\
[M_{\pm 1}, P_{\mp} ] & = i \eta_{+ -} P_1\,, \label{pc2}
\end{align}
where $\eta_{+ -}=\eta_{- +}$ and $\eta_{1 1}$ denote Lorentzian metric components (cf. \cite{BP2}),
we can find that for each $n=1,2,\dots \in \mathbb{N}$ there is a $\dagger$- preserving embedding of \eqref{pc1}-\eqref{pc2} to the infinite-dimensional Lie algebra \eqref{bms3d}
\begin{align}\label{pc3}
M_{+-} & =-  i l_0\,, \quad M_{\pm 1} = \pm \frac{i }{\sqrt{2}}l_{\pm n}\,, \\
P_1 & = -i T_0, \quad P_{\pm} = \frac{i}{\sqrt{2}} T_{\pm n}\,,
\label{pc4}\end{align}
if one identifies $\eta_{+ -}=\eta_{- +}=-\eta_{1 1}=n$. In the formula above we rescaled the metric instead of using the embedding with rescaled generators, because the rescaling of $l_0$ changes the algebra \eqref{bms3d} to which we embed.
Note that, just as in the 4d case, these isomorphic embeddings are not automorphisms,
since the metric components depend on $n$ \footnote{This property makes it different from a phenomenological algebra introduced recently in \cite{Gomis} for the description of curvature corrections to the isometry algebra of flat space-time. Another difference is that in our case all embeddings have two common elements $l_0$ and $T_0$.}. They are
not compatible with automorphisms of the full $\mathcal B^3$ algebra as explained in Appendix A.
Each Poincar\'e sector has its own mass Casimir $\cC_n\equiv
\eta^{ab}P_aP_b= 2T_{-n}T_{+n}-T^2_0$, which however is not  a central element for the entire $\mathcal B^3$ (BMS Lie algebra is centreless and admits central extension with two central charges).  Of course, the identification \eqref{pc3} - \eqref{pc4} is unique up to an automorphism of the full $\mathcal B^3$  algebra; e.g. one can apply an involutive automorphism: $l_m\mapsto-l_{-m}, T_m\mapsto -T_{-m}$\,.
We can therefore see that $\mathcal B^3$ can be composed entirely from subalgebras which are isomorphic to the  Poincar\'e algebra. Moreover, it turns out that they are maximal finite-dimensional subalgebras of \eqref{bms3d}.

Let us construct explicitly the vacua corresponding to different choices of the Poincar\'e subgroup. We take as a starting point the general solution of the vacuum Einstein equations in Bondi gauge \cite{Barnich:2010eb}
\begin{equation}
	d s^{2}=\Theta(\phi) d u^{2}-2 d u d r+\left(u \Theta^{\prime}(\phi)+\Xi(\phi)\right) d u d \phi+r^{2} d \phi^{2},
\end{equation}
where $\Theta$ and $\Xi$ are arbitrary periodic functions of $\phi$. In \cite{Barnich:2010eb} a further generalization is considered by allowing metrics with angular part $r^2e^{2\varphi}d\phi^2,\; \varphi=\varphi(u,\phi)$ but we consider here only the case $\varphi=0$.
Writing this solution using $z=e^{i\phi}$ one obtains
\begin{align}\label{solutionL}
	ds^2=\Theta(z) du^2-2dudr+(u\Theta(z)'-i\Xi(z) z^{-1})dudz-\frac{r^2}{z^2}dz^2.
\end{align}
Plugging the metric components into \eqref{3dsupertran} and \eqref{3dsuperrot} one obtains for the generators of supertranslations
\begin{equation}
	T_n=z^n\partial_u-\left(n^2z^n-\frac{nz^{n+1}}{r}g_{zu}\right)\partial_r+\frac{nz^{n+1}}{r}\partial_z
\end{equation}
and of superrotations
\begin{align}
	l_n=iunz^n\partial_u-i\left(nrz^n+un^3z^n+\frac{un^2z^{n+1}}{r}g_{uz}\right)\partial_r+iz^{n+1}\left(1+\frac {un^2}{r} \right)\partial_z.
\end{align}

By demanding that the Lie derivative of \eqref{solutionL} with respect to the generators comprising the embeddings vanishes we obtain a metric that is invariant under the action of these generators. One finds that the only non-vanishing components of the Lie derivative with respect to $T_n$ and $l_n$ are
\begin{align}
	\mathcal L_{T_n} g_{uu}=-\frac nr z^{n+1}\Theta',\quad \mathcal L_{T_n}g_{uz}=z^{n-1}(\Theta n+n^3) \nonumber\\
	\mathcal L_{l_n} g_{ur}=-\frac{2un^2z^{n+1}}{r^2}g_{zu},\quad \mathcal L_{l_n}=4un^2z^{n-1}g_{zu}.
\end{align}
All Lie derivatives vanish for $n=0$ and for arbitrary $n$ if we demand $\Xi=\Theta'=0$ and $\Theta=-n^2$. Notice that for $\Theta=-1$ we correctly recover invariance only under standard Poincar\'e transformations with $n=0,\pm 1$. Since the metric is also invariant under the action of $T_{-n},l_{-n}$ using the same demands on $\Theta,\Xi$ we can conclude that the generators $T_0, T_{\pm n}, l_0, l_{\pm n}$ are the exact Killing vectors of the following solution of Einstein vacuum equations
\begin{equation}
	ds^2=-n^2du^2-2dudr-\frac{r^2}{z^2}dz^2.
\end{equation}
This metric can be diagonalized to become
$$
ds^2 = - dt^2 +\frac1{n^2}\, dr^2 + r^2 d\phi^2\,,
$$
which after rescaling $r\mapsto r/n$ becomes
\begin{equation}\label{conical}
  ds^2 = - dt^2 + dr^2 + r^2 n^2 d\phi^2\,,
\end{equation}
For $n=1$ this is the standard flat space metric, but for $n\neq1$ \eqref{conical} is the metric of the space with conical singularity. Notice %\textcolor{red}{
	that this connection between the solutions with a  conical defect and the Poincar\'e subgroups has previously been noted in \cite{Barnich:2015uva}  and that a similar singular nature of asymptotic vacua in 4d was noticed in \cite{Compere:2016jwb}. We will further discuss the vacua in three and four dimensions in a forthcoming publication and now turn to the major theme of this paper, the deformation of the BMS algebra.

\section{Quantum deformations from 3d BMS algebra}\label{sec3}

\subsection{Quantum groups}

After discussing the structure of BMS algebra in 3 and 4 dimensions, let us now turn to the discussion of their deformations. As it was in the case of our previous investigations in \cite{Borowiec:2018rbr} in deforming these algebras, their Poincar\'e subalgebras will play a special role.

In the rest of the paper we will use the following notation. The BMS Lie algebra we already introduced, $\mathcal B_{\text{ext}}^4$ in four dimensions, is denoted by $\mathcal{B}^{3/4}$ in three and four dimensions. Its universal enveloping algebra, consisting of polynomials in the generators subject only to the Lie algebra commutation relations, is called $U \mathcal{B}^{3/4}$. By the very definition it contains the initial Lie algebra as a subspace spanned by the generators. This enveloping algebra can be endowed with a (primitive on generators $X$ of the Lie algebra) coproduct $\Delta_0$, counit $\varepsilon$ and antipode $S_0$ to form a Hopf algebra
\begin{align}\label{prim}
\Delta_0(X) = X \otimes 1 + 1 \otimes X, \quad S_0(X) = -X, \quad \varepsilon(X) =0.
\end{align}

 Such standard Hopf algebra structure will be then quantized by using Drinfeld's twist deformation techniques \cite{Drinfeld:1983rx,Drinfeld2}. To this end one has to extend  $U \mathcal{B}^{3/4}$ by introducing a new commuting generator denoted as ${1/\kappa}$\footnote{ For the sake of physical applications $\kappa$ has mass dimension and frequently is identified with Planck mass.}. Strictly speaking, one needs a topological extension of the enveloping algebra  which allows for formal power series in ${1}/{ \kappa}$ (denoted by $U\mathcal{B}^{3/4}[[1/\kappa]]$). A new, deformed, Hopf algebra equipped with a twist-deformed coproduct $\Delta_{n, \epsilon}$  and a compatible antipode $S_{n, \epsilon}$ becomes according to Drinfeld's terminology a `quantum group' which we will call $U\mathcal B_{n, \epsilon}^{3/4}$ for brevity\footnote{The indices stand for a particular twist discussed below.}.
 The coproduct in our case will be obtained by a similarity transformation
 \begin{align}\label{tw}
\Delta_{n, \epsilon}(X) = \mathcal{F}_{n, \epsilon} \Delta_0(X) \mathcal{F}_{n, \epsilon}^{-1}
\end{align}
 with a twist $\mathcal{F}_{n, \epsilon} \equiv a_{\alpha} \otimes b^{\alpha}$ that has to satisfy two-cocycle conditions
 \begin{align}
\mathcal{F}_{12} \dot (\Delta_0 \otimes 1) (\mathcal{F}) = \mathcal{F}_{23} \dot (1 \otimes \Delta_0) (\mathcal{F}), \quad \varepsilon (a_{\alpha}) b^{\alpha} = 1,
\end{align}
where $\mathcal{F}_{12} = a_{\alpha} \otimes b^{\alpha} \otimes 1$ etc.

For every Hopf algebra which is obtained as a quantum deformation of a Lie algebra (so-called quantized universal enveloping algebra) one introduces a  formal  variable enabling one to expand deformed coproduct and antipode. In our case it is a dimension full parameter $1/\kappa$ from which one can calculate the first order term of the antisymmetrized coproduct.
It turns out that for each coproduct this so called `classical limit' applied to the Lie algebra generators $X$
\begin{equation}\label{limit}
\delta_{n, \epsilon}(X)= \lim_{\kappa\rightarrow \infty}\kappa (\Delta_{n, \epsilon}(X)-\Delta_{n, \epsilon}^{21}(X))
\end{equation}
defines on the initial Lie algebra  $\mathcal{B}^{3/4}$ the cobracket $\delta_{n, \epsilon}$ providing the structure of a Lie bialgebra denoted as $\mathcal{B}^{3/4}_{r_{n, \epsilon}}$. Moreover, such Lie bialgebra structure is implemented by an anitsymmetric element $r_{n, \epsilon}$, the so-called classical $r_{n, \epsilon}$-matrix satisfying the classical Yang-Baxter equation (see below).
This means we obtain a  cobracket applied to an arbitrary element $X$ of the initial Lie algebra in the form
\begin{align}\label{cobra}
\delta_r(X) = ad_{r} \triangleright X\equiv [r, X \otimes 1 + 1 \otimes X]
\end{align}
 This element determines a classical Lie bialgebra structure on the Lie algebra in question that in turn undergoes the process of Hopf algebraic quantization.

 $\Delta^{21}$ in eq.\eqref{limit} denotes the flipped (co-opposite) coproduct which for coboundary deformations is provided by the quantum $r$-matrix $\mathcal R$: $\Delta^{21}=\mathcal R\Delta\mathcal R^{-1}$. In this way the formula \eqref{limit}  expresses on the one hand the so-called classical limit of quantum deformations and on  the other hand relations between classical and quantum $r$-matrices, i.e. the relations  between  solutions of classical and quantum Yang-Baxter equations \cite{Drinfeld:1983rx}. Furthermore, the quantum $r$-matrix is related to the twist via $\mathcal R = \mathcal{F}_{21} \mathcal{F}^{-1}$.

As stated above the $r$-matrix has to fulfill the classcial Yang-Baxter equation
\begin{align}\label{omega}
[[r, r]] = \Omega,
\end{align}
where $[[ , ]]$ denotes a Schouten bracket, and $\Omega$ is an ad-invariant element.
Such deformations are called coboundary; if the rhs is zero then they are  triangular.
In such a case we call the corresponding bialgebra structure coboundary triangular.
All triangular deformations are derivable  by a twisting procedure (like eq.\eqref{tw}) as described e.g. in \cite{Aschieri:2017ost,BP2,Borowiec:2018rbr} and references therein.

The physically  important example of Hopf algebra that we will use as a starting point here is $\kappa$-Poincar\'e Hopf algebra, which exists in three different versions characterized by the fixed vector $\tau$. This vector is spacelike, timelike or lightlike,  and its role can be seen most easily by considering the three dual algebras\footnote{We will see in the next section how the algebra of the duals is related to the coproduct of the Hopf algebra.}, the $\kappa$-Minkowski algebra
\begin{align}
[x^{\mu}, x^{\nu}] = \frac{i}{\kappa} (\tau^{\mu} x^{\nu} - \tau^{\nu}x^{\mu}).
\end{align}
Furthermore it is known that out of these three different $\kappa$-Poincar\'e versions, only the lightlike one is coboundary triangular and can thus be constructed by a twist.

The particular variant of deformation is therefore chosen to be corresponding to the lightlike (also called lightcone) $\kappa$-Poincar\'e because it automatically guarantees that the embeddings of the Poincar\'e sub-Hopf algebras are consistent in the sense that the new coproduct is a homomorphism for all $X, X' \in U\mathcal{B}_{n, \epsilon}$
\begin{align}
\Delta (XX') = \mathcal{F} \Delta_0(XX') \mathcal{F}^{-1}= \mathcal{F}\Delta_0(X) \mathcal{F}^{-1} \mathcal{F} \Delta_0(X') \mathcal{F}^{-1} = \Delta(X) \Delta(X') .
\end{align}
In fact, one can even make a stronger argument that only the coboundary Hopf algebra is possible in $U\mathcal{B}$. In the Poincar\'e algebra there is only one nontrivial candidate for the ad-invariant element in eq.\eqref{omega}
\begin{align}
\Omega = M_{\mu \nu} \wedge P^{\mu} \wedge P^{\nu}
\end{align}
but this is not ad-invariant in the entire BMS. Thus the Schouten bracket has to vanish which is the defining relation for a coboundary Hopf algebra.

Another reason to choose the lightlike incarnation is that it can be naturally formulated in lightcone coordinates
\begin{align}
x^+ = \frac{x^0 + x^2}{\sqrt{2}}\,, \quad x^- = \frac{x^0-x^2}{\sqrt{2}}\,,\quad x^1 = x^1
\end{align}
suited to the supertranslations acting on null surfaces.

To explicitly construct the lightcone $\kappa$-Poincar\'e one can first canonically assign a Hopf algebra structure to $U \mathcal{B}^3$ by using the undeformed coproduct and antipode defined in eq.\eqref{prim}
for every generator $X$.
This trivial coalgebra structure then has to be deformed as it was done in \cite{Borowiec:2018rbr} in the 4d case by using a 2-cocycle twist. There the whole Hopf algebra structure was obtained and used to argue that the coproduct ``mixes'' the elements of the Poincar\'e algebra and the other BMS generators.

Now the main difference is that the different embeddings of the Poincar\'e algebra into $\mathcal{B}^3$  allow for twists that turn out to lead to non-isomorphic Hopf algebras living in the same universal enveloping algebra structure \footnote{In four dimensions similar embeddings exist and we start with the three dimensional case mainly for simplicity.}.
This means there is a family of unitary twisting elements $\mathcal{F}_{n, \epsilon}$ satisfying two-cocycle and normalization conditions which read \footnote{Unitary twist provides a real deformation.}
\begin{align}
\mathcal{F}_{n,\epsilon} & = \exp \left( - \epsilon \eta^{11}\frac{i}{\kappa} M_{+ 1} \otimes P_1 \right) \exp \left(- i M_{+-} \otimes \eta^{+-}\log \left( 1 + \frac{P_+}{\kappa} \right)  \right) \\
& = \exp \left( -\epsilon\frac{\imath}{n\kappa \sqrt{2}} l_n \otimes T_0 \right) \exp \left(-l_0 \otimes 1/n\log \left( 1 - \frac{\imath}{\kappa \sqrt{2}}T_n \right) \right),
\end{align}
where $\epsilon= 0,1$. We introduced the factor $\epsilon$ which can be either 1 corresponding to the full deformation or 0 which leaves only the Jordanian part of the twist. It should be stressed that the Jordanian twist is physically viable on its own and also corresponds to a Lie-algebra type deformation of the Minkowski sector as we will show below.

Using the Hadamard formula
$$
e^A B e^{-A} = \sum_{n=0}^{\infty} \frac{1}{n!} \underset{\text{n times}}{[\underbrace{A, [A, ...[A, }} B]...]
$$
we can calculate the deformed coproducts by
\begin{align}
\Delta_{n,\epsilon} (l_m) = \mathcal{F}_{n, \epsilon} \Delta_0(l_m) \mathcal{F}_{n, \epsilon}^{-1}\,, \quad
\Delta_{n, \epsilon} (T_m) = \mathcal{F}_{n, \epsilon} \Delta_0 (T_m) \mathcal{F}_{n, \epsilon}^{-1}
\end{align}
and we obtain
\begin{align}
\Delta_{n,\epsilon} (l_m) =& 1  \otimes l_m +i \frac{(m-n)}{n \sqrt{2}\kappa} l_0 \otimes T_{m+n} \Pi_{+n}^{-1} - i \frac{m}{n\sqrt{2}\kappa} l_n \otimes T_m + \epsilon \frac{(m-n)}{n 2 \kappa^2} l_n \otimes T_0 T_{m+n} \Pi_{+n}^{-1} \nonumber \\
&+  l_m \otimes \Pi^{-\frac{m}{n}}_{+n} + \epsilon \sum_{k=1}^{\infty} \left( \frac{i}{n\sqrt{2}\kappa} \right)^k \frac{1}{k!} f^k_{n m} l_{m+kn} \otimes T_0^{k} \Pi_ {+n}^{-\frac{m}{n}}, \label{copr1}\\
\Delta_{n,\epsilon} (T_m) = & 1 \otimes T_m + T_m \otimes \Pi_ {+n}^{-\frac{m}{n}} +  \epsilon \sum_{k=1}^{\infty}\left( \frac{i}{n\sqrt{2}\kappa} \right)^k \frac{1}{k!} f^k_{nm} T_{m+kn} \otimes T_0^k \Pi_{+n}^{-\frac{m}{n}}, \label{copr2}
 \end{align}
where
\begin{align}
f^k_{nm} =  \prod^{k-1}_{j=0} (n-(m+jn)), \quad \Pi_{+n} = \left( 1 - i \frac{T_n}{\sqrt{2} \kappa}\right).
\end{align}
Each coproduct label by $n=1, 2, \ldots$ represents a different Hopf algebra deformation of the enveloping algebra \footnote{In fact, one needs a topological extension of the enveloping algebra which allows for formal power series in $\frac{1}{ \kappa}$.} a.k.a quantum group or more precisely a quantum enveloping algebra. Moreover, due to the value of $\epsilon=0$ or $\epsilon=1$ one has two quantum group structures for each $n$: Jordanian or extended Jordanian. Furthermore, each of these quantum groups admits only one $\kappa$-Poincar\'e as Hopf subalgebra.

%As a bialgebra the three dimensional $\kappa$-Poincar\'e is a subbialgebra of the BMS since it is a subalgebra and a coideal like in the four dimensional case.

Each quantum group has as a classical limit the corresponding Lie bialgebra structure. In our case
the bialgebra cobrackets \eqref{cobra} are controlled ($\delta_{r} (X) = ad_{r} \triangleright X$)
by classical $r$-matrices in the form
\begin{align}\label{cl_r_matrix}
%\delta_{r_n} (X) = ad_{r_n} \triangleright X, \quad
r_{n,\epsilon} =\eta^{- +} M_{+ -} \wedge P_+ + \epsilon\, \eta^{1 1} M_{+1} \wedge P_1 = \frac{1}{\sqrt{2}n} \left( l_0 \wedge T_n + \epsilon\, l_n \wedge T_0 \right)
\end{align}
and explicitly reads
\begin{align}
\delta_{n,\epsilon} (l_m) = & \frac{i}{\sqrt{2}n} \big( (n-m) l_0 \wedge T_{m+n} - \epsilon m l_n \wedge T_m \nonumber \\
&- \epsilon (m-n)) l_{m+n} \wedge T_0 -m l_m \wedge T_n \big) \,,\\
\delta_{n,\epsilon} (T_m) =& \frac{i}{\sqrt{2}n}( \epsilon (n-m) T_{m+n} \wedge T_0 -m T_m \wedge T_n )\,.
\end{align}
This bialgebra structure will be needed to construct a dual algebra that can be interpreted as a generalized coordinate space.

\subsection{Bialgebra duals for BMS in 3d gravity}

It is well-known that the dual of finite-dimensional Lie bialgebra $(\mathfrak{g}, [,],\delta)$ is a Lie bialgebra structure determined on the dual vector space with a cobracket dual to the Lie bracket and vice-versa, i.e. $(\mathfrak{g}^*, [,]_\circ=\delta^*,\delta^\circ=[,]^*)$. Usually, the dual cobracket is not a coboundary and the dual Lie algebra structure is solvable. However, for the infinite-dimensional case we are dealing with here the situation is much more complicated, see Appendix B.

Here, the dual space is constructed using the expression of $\mathcal{B}^3$ in terms of the abstract variable $x$, its formal inverse $x^{-1}$ and their respective duals (see Appendix B for details). It is spanned by elements  $\{(\zeta^j, \chi^k), j, k \in \mathbb{Z} \}$ and will be denoted as $\mathcal{B}^{3 *}_{r_{n,\epsilon}}$. \footnote{For the sake of simplicity, following \cite{Song:2019}, we are using a pseudo-basis $\zeta^m,\chi^k; m,k\in \mathbb{Z}$ in $\mathcal{B}^{3 *}$ instead of true basis $\zeta^{a,i}, \chi^{b,j}$ in $\mathcal{B}^{3 \circ}$ (see Appendix B ).}

\par
Turning now to the structure on the dual space (for definitions see Appendix  \ref{AppendixB}) we use that the cobracket, which is dual to the Lie algebra bracket \eqref{bms3d}, is calculated by
\begin{align}
\braket{ \delta^\circ(\zeta^m), l_p \otimes l_q} &= \braket{\zeta^m, [l_p, l_q]}, \\
\delta^\circ(\zeta^m) &= \sum_{i+j = m} (j-i) \zeta^i \otimes \zeta^j , \label{d1}\\
 \braket{ \delta^\circ(\chi^m), l_p \otimes T_q} &= \braket{\chi^m, [l_p, T_q]}
 \end{align}
 and we obtain
 \begin{align}\label{d2}
 \delta^\circ(\chi^m) = \sum_{i+j = m} (j-i) (\zeta^i \otimes \chi^j + \chi^i \otimes \zeta^j) .
\end{align}
In fact the formulas  above are not well-posted since the sums in \eqref{d1}, \eqref{d2} are infinite. This is because the eq. $i+j=m$ for fixed $m\in\mathbb{Z}$ has infinitely many solutions. However, it becomes finite if we put the constraints $i, j, m\geq -1$, i.e. for the case of the one-sided $\cW(2,2)_1$ algebra (cf. Appendix B).

For the brackets which are dual to the bialgebra cobrackets $\delta_{n,\epsilon}$ we find in a similar way
\begin{align}
\braket{[\zeta^k, \zeta^j]_{n,\epsilon}, l_m} &= \braket{\zeta^k \otimes \zeta^j, \delta_{n,\epsilon}(l_m)}, \\
[\zeta^k, \zeta^j]_{n,\epsilon} &= 0, \\
[\zeta^k, \chi^j]_{n,\epsilon} &= \frac{i}{\sqrt{2}n}((2n -j) \delta_{k, 0} \zeta^{j-n} + k\delta_{j, n} \zeta^k) - \epsilon (k \Leftrightarrow j), \\
[\chi^k, \chi^j]_{n,\epsilon} &= \frac{i}{\sqrt{2}n}(-k\delta_{j, n} \chi^k + \epsilon (2n -k) \delta_{j, 0} \chi^{k-n}) - (k \Leftrightarrow j).
\end{align}
Particularly, for each embedding labeled by $n=1,2\ldots$, the duals of Poincar\'e translation generators are defined such that $ \braket{P_{\mu}, x^{\nu}} = \delta^{\nu}_{\mu}$ and read
\begin{align}\label{pcduals}
x^{\pm} \equiv \pm i \sqrt{2} \chi^{\pm n}, \quad x^1 = i \chi^0.
\end{align}
Thus we obtain
\begin{align}
[x^+, x^-]_{n, \epsilon} = \frac{1}{\kappa} x^-, \quad [x^+, x^1]_{n, \epsilon} = \epsilon \frac{1}{\kappa} x^1, \\
[x^-, x^1]_{n, \epsilon} = - \epsilon \frac{3i}{\kappa }  \chi^{-2n},
\end{align}
which mimics the three dimensional $\kappa$-Minkowski spacetime algebra but is not a subalgebra for the full deformation. The smallest subalgebra containing the $\kappa$-Minkowski is given by $\{ \chi^{-(k-1)n}, k \in \mathbb{N} \}$ and to obtain a coideal one has to include all $\chi^m$. This result is not unexpected in the sense that the dual of an algebra with a subalgebra $\mathcal{S}$ does not necessarily contain a subalgebra that is dual to $\mathcal{S}$. Physically speaking this could express that in asymptotically flat spacetimes it is only possible to define local coordinates that are constructed by dualizing a given embedding (i.e. the Poincar\'e algebra). Coordinates that are dual to the entire deformed BMS bialgebra $\mathcal{B}^3_{r_{n, 1}}$ on the other hand do not form a closed $\kappa$-Minkowski subalgebra.
For the Jordanian twist this problem does not appear as the algebra closes. In this case it is therefore possible to define coordinates dual to $\mathcal{B}^3_{r_{n, 0}}$.

\subsection{Quantum Lie algebra generators}

In the present section we employ the quantum Lie algebra formalism. Following \cite{Wess06}  (for recent applications see also \cite{Aschieri:2017ost,Kuznetsova:2018ncv,Fiore2020}) we define the twisted (quantum Lie algebra) generators of $U\mathcal{B}^3_{n}$ as
\begin{align}\label{defalg}
l^{\mathcal{F}}_m \equiv \mathcal{D}^{\mathcal{F}}(l_m) = (\bar f^{\alpha} \triangleright l_m) \bar f_{\alpha}
\end{align}
for the twist $\mathcal{F}$ with inverse $\mathcal{F}^{-1} \equiv \bar f^{\alpha} \otimes \bar f_{\alpha}$.
A quantum Lie algebra $g^{\mathcal{F}}$ corresponds to a quantum universal envelope $U g^{\mathcal{F}}$ (endowed with a twisted Hopf algebra structure) in analogy to the classical case where there is a one-to-one correspondence between a Lie algebra $g$ and a universal enveloping algebra $U g$. $g^{\mathcal{F}}$ has to fulfill three conditions
\begin{align*}
&i) \quad  g^{\mathcal{F}} \;  \text{generates}\;  Ug^{ } \\
&ii) \quad  \Delta^{\mathcal{F}} (g^{\mathcal{F}}) \subset g^{\mathcal{F}} \otimes 1 + U g^{\mathcal{F}} \otimes g^{\mathcal{F}} \\
&iii) \quad  [ g^{\mathcal{F}}, g^{\mathcal{F}}]_{\mathcal{F}} \subset g^{\mathcal{F}}
\end{align*}
where the bracket is defined as the adjoint action with the twisted coproduct and antipode.
In \cite{Aschieri:2017ost} it was argued that the elements of the quantum Lie algebras are to be interpreted as the physical momenta (or supermomenta in our case). Thus also other phenomenologically relevant relations like the dispersion relations dictated by the form of the Casimir operator and the addition of (super)momenta can be calculated.
\par
The action in eq.\eqref{defalg} is just the standard adjoint action (i.e. defined with the undeformed coproduct) and so we have that the Casimir $\mathcal{C}_n \equiv P_{n \mu} P_n^{\mu}$ of the Poincar\'e subalgebra corresponding to a given twist $\mathcal{F}_n$ is undeformed
\begin{align}
\mathcal{C}^{\mathcal{F}_n}_n = P_{n \mu}^{\mathcal{F}_n}P_{n}^{\mathcal{F}_n \mu} =  \mathcal{D}^{\mathcal{F}_n} ( \mathcal{C}_n) = \mathcal{C}_n = 2 P_{n+} P_{n-} - P_{n1}^2
\end{align}
where we used that $\mathcal{D}^{\mathcal{F}}$ is a homomorphism and the adjoint action on the Casimir vanishes except for the unit.
\newline
If only the Jordanian part $\mathcal{F}_n^J \equiv \mathcal{F}_n(\epsilon = 0)$ of the twist $\mathcal{F}_n$ is considered one finds
\begin{align}
\mathcal{D}^{\mathcal{F}_n^J} (P_{m +}) &= P_{m +} (1 + \frac{P_{n+}}{\kappa})^{-\frac{m}{n}}\\
\mathcal{D}^{\mathcal{F}_n^J} (P_{m -}) &= P_{m -} (1 + \frac{P_{n+}}{\kappa})^{\frac{m}{n}}\\
\mathcal{D}^{\mathcal{F}_n^J} (P_{m 1}) &= P_{m 1}
\end{align}
so that the Casimir is also undeformed. Note that this coincides with the results of \cite{Kuznetsova:2018ncv} for $m=n$. In the same way one can also calculate the twisted generators in the superrotation sector
\begin{align}\label{defl}
\mathcal{D}^{\mathcal{F}_n^J}(l_m) = l_m \left(\Pi_{+n}\right)^{- \frac{m}{n}}.
\end{align}
The full twist is more complicated and so we calculate in first order $1/\kappa$
\begin{align}
\mathcal{D}^{\mathcal{F}_n}(P_{m+}) &\approx P_{m+} + \frac{1}{2 n \kappa} ((n-m) T_{n+m} T_0- m T_m T_n) \\
\mathcal{D}^{\mathcal{F}_n}(P_{m-}) &\approx P_{m-} + \frac{1}{2 n \kappa} (-(n+m) T_{n-m} T_0- m T_{-m} T_n )\\
\mathcal{D}^{\mathcal{F}_n}(P_{m1}) &\approx P_{m1} - \frac{T_m T_0}{\sqrt{2}  \kappa}
\end{align}
so that
\begin{align}
\mathcal{D}^{\mathcal{F}_n}(2 P_{m+} P_{m-} - P_{m1}^2) & \approx  2  P_{m+} P_{m-} - P_{m1}^2 + \frac{2 P_{m+}}{n \kappa} \left( \frac{(n+m)}{2} iT_{n-m} P_{m1} - \frac{m}{n}P_{m-} P_{n+}\right) \nonumber \\
& \phantom{mn} \frac{2 P_{m-}}{n \kappa} \left( \frac{-(n+m)}{2} iT_{n+m} P_{m1} + \frac{m}{n}P_{m+} P_{n+}\right) + \frac{2 P^2_{m1}}{\kappa} P_{m+} .
\end{align}
meaning that the full $\kappa$-Poincar\'e deformation changes the Casimir of any other embedding already at first order. This change is not purely multiplicative, i.e. $\mathcal{C}^{\mathcal{F}_n}_m$ is not proportional to $\mathcal{C}_m$, which means that both the dispersion relation of massless and massive particles is affected.
\par
Since the coproduct induces a representation on multiparticle states one can infer the total (super)momentum from it. We make use of the general formula
\begin{align}\label{twcop}
\Delta_{\mathcal{F}}(T^{\mathcal{F}}) = T^{\mathcal{F}} \otimes 1  + \bar R^{\alpha} \otimes (\bar R_{\alpha}(T) )^{\mathcal{F}}
\end{align}
with the inverse quantum R-matrix $R^{-1} = \mathcal{F} \mathcal{F}_{21}^{-1} \equiv \bar R^{\alpha} \otimes \bar R_{\alpha}$.
Considering only the Jordanian part the coproduct reads
\begin{align}
\Delta_{n} (T_m^{\mathcal{F}^J_n}) = T_m^{\mathcal{F}^J_n} \otimes 1 + e^{- \frac{l^{\mathcal{F}_n^J}_0}{n}}  \frac{1}{{\left(\Pi_{+n}\right)}^{\frac{m}{n}}} \otimes T_m^{\mathcal{F}^J_n},
\end{align}
where $\Pi_{+n}(P_{+n})$ can be expressed by twist deformed generators via
\begin{align}
P_{+n}  = \frac{P_{+n}^{\mathcal{F}_n^J}}{1 -\frac{P_{+n}^{\mathcal{F}_n^J}}{\kappa }}.
\end{align}
For $\epsilon = 1$ we find in $\mathcal{O}(1/\kappa)$
\begin{align}
\Delta_{n}  (T_m^{\mathcal{F}_n}) = T_m^{\mathcal{F}_n} \otimes 1 + 1 \otimes T_m^{\mathcal{F}_n} + \frac{i}{\sqrt{2} \kappa} \left(-\frac{m}{n}T_{n} \otimes T_m^{\mathcal{F}_n} + \epsilon \frac{n-m}{n} T_0 \otimes T_{m+n}^{\mathcal{F}_n} \right),
\end{align}
where $T_0$ is expressed by inverting the relation
\begin{align}
T_0^{\mathcal{F}_n} = T_0 \left( 1 - \frac{T_n}{\Pi_{+n}}\frac{\epsilon i}{\sqrt{2}\kappa}\right).
\end{align}

With the help of
\begin{align}\label{fbrak}
[l^{\mathcal{F}}, T^{\mathcal{F}}]_{\mathcal{F}} = l^{\mathcal{F}} T^{\mathcal{F}} - (\bar R^{\alpha}(T))^{\mathcal{F}} (\bar R_{\alpha}(l))^{\mathcal{F}}
\end{align}
one can calculate the deformed bracket. Here, the inverse R-matrix for $\epsilon = 0$ is given by
\begin{align}
R^{-1}_n = \mathcal{F}_n\mathcal{F}^{-1}_{n 21} = \exp \left( \frac{-l_0}{n} \otimes \log \left(1+ \frac{P_{+ n}}{\kappa} \right) \right) \exp \left( \log \left(1 + \frac{P_{+n}}{\kappa} \right) \otimes \frac{l_0}{n} \right)
\end{align}
and we use
\begin{align}
\mathcal{F}^{-1}_n = \sum_{j =0}^{\infty} \frac{1}{j !} \left(\frac{l_0}{n} \right)^{\underline{j}} \otimes \left( \frac{P_+}{\kappa}\right)^j
\end{align}
with $X^{\underline{j}} = X (X-1) ... (X-(j-1))$ the lower factorial.

As a result, eq.\eqref{fbrak} yields
\begin{align}
[l_m^{\mathcal{F}^J_n}, l_p^{\mathcal{F}^J_n}]_{\mathcal{F}^J_n} = & l_m^{\mathcal{F}^J_n} l_p^{\mathcal{F}^J_n} +  l_p^{\mathcal{F}^J_n} l_m^{\mathcal{F}^J_n}+ \frac{1}{2\kappa^2} \frac{(n+p)(n-p)(n-m) m}{n^2} T_{n+ p}^{\mathcal{F}^J_n} T_{n+m}^{\mathcal{F}^J_n} \nonumber \\
& + \frac{-i}{\sqrt{2}\kappa n} \left( p(n-m) l_p^{\mathcal{F}^J_n} T_{n+m}^{\mathcal{F}^J_n} - m(n-p) T_{n+p}^{\mathcal{F}^J_n} l_m^{\mathcal{F}^J_n} \right), \label{defcom1}
\end{align}
and, together with eq.\eqref{defl} and \eqref{bms3d}, one has
\begin{align}
[l_m^{\mathcal{F}^J_n}, l_p^{\mathcal{F}^J_n}]_{\mathcal{F}^J_n}  = (m-p) l_{m+p}^{\mathcal{F}^J_n} + \frac{i}{\sqrt{2}\kappa} \frac{(m-(n+p))(n-p) m}{n^2} T_{n+ m+ p}^{\mathcal{F}^J_n}. \label{res}
\end{align}
Apparently the structure constants are not identical to the undeformed algebra relations already in the Jordanian case. However for
\begin{align}
[l_m^{\mathcal{F}^J_n}, T_p^{\mathcal{F}^J_n}]_{\mathcal{F}^J_n} = l_m^{\mathcal{F}^J_n} T_p^{\mathcal{F}^J_n} - T_p^{\mathcal{F}^J_n} l_m^{\mathcal{F}^J_n} + \frac{i}{\sqrt{2}\kappa}(n-m) \left( - \frac{p}{n} \right) T_{n+m}^{\mathcal{F}^J_n}T_p^{\mathcal{F}^J_n}
\end{align}
we find
\begin{align}
[l_m^{\mathcal{F}^J_n}, T_p^{\mathcal{F}^J_n}]_{\mathcal{F}^J_n} = (m-p)  T_{p+m}^{\mathcal{F}^J_n}, \label{res2}
\end{align}
i.e. these structure constants are undeformed.
\par To summarize, it is noteworthy that there are physical predictions which allow a distinction between undeformed, Jordanian and $\kappa$-Poincar\'e twisted structures. Both versions of the twist predict an altered Leibnitz rule, i.e. a deformed prescription of how supermomenta are added. Furthermore the $\kappa$-Poincar\'e twist deforms the energy-momentum dispersion relation for supermomenta such that also massive particles are affected whereas this relation is undeformed for the Jordanian twist for all scalar particles.
\newline
As stated earlier for both the Jordanian and extended Jordanian twist there are infinitely many different quantum groups labeled by $n \geq 1$ and we can now see that the coproducts dictating the addition of physical momenta explicitly depend on the embedding. The extra structure constants for the quantum Lie algebra we found also contain this explicit dependence for both types of twist and for the extended Jordanian twist even the dispersion relation varies with different $n$. In the next section we therefore investigate a twist that does not single out a certain embedding.

\section{Abelian twist}\label{sec5}

In section \ref{sec2} we noted that there are infinitely many embeddings of the Poincar\'e algebra into the BMS algebra, which however overlap since they all contain $l_0, T_0$. Thus, in order to construct an embedding-independent twist one could use these elements. The $r$-matrix is given by
\begin{align}\label{ab-r}
r_A = \frac{i}{\kappa}l_0 \wedge T_0
\end{align}
and the corresponding twist reads
\begin{align}
\mathcal{F}_A = \exp \left( \frac{i}{\kappa} l_0 \otimes T_0 \right).
\end{align}
It is an abelian twist since $l_0$ and $T_0$ commute and the $r$-matrix automatically satisfies the classical Yang-Baxter equation \eqref{omega} with a vanishing rhs.
\par
The Hopf-algebra structure associated with the twist is calculated analogously to the lightcone $\kappa$-Poincar\'e twist and we find
\begin{align}
\Delta_{A}(l_m) = & \mathcal{F}_A \Delta_0 (l_m) \mathcal{F}_A^{-1} \nonumber \\
=& 1 \otimes l_m - \frac{im }{\kappa} l_0 \otimes T_m + l_m \otimes \exp \left(- \frac{im}{\kappa} T_0 \right) \\
\Delta_A(T_m) = & 1 \otimes T_m + T_m \otimes \exp \left(- \frac{im}{\kappa} T_0 \right).
\end{align}
The cobracket defined with \eqref{ab-r}
\begin{align}
\delta_{r_A} (l_m) = \frac{-im}{\kappa} (l_0 \wedge T_m + l_m \wedge T_0),
\delta_{r_A} (T_m) = \frac{-im}{\kappa} (T_m \wedge T_0)
\end{align}
turns $\mathcal{B}^3$ into a Lie-Bialgebra $\mathcal{B}^3_{r_A}$ and by dualising this structure one obtains
\begin{align}
[\zeta^k, \chi^j]_A = - \frac{i}{\kappa} ( j \delta_{k 0} \zeta^j + k \delta_{j 0} \zeta^k) , \\
[\chi^k, \chi^j]_A =  - \frac{i}{\kappa} ( k \delta_{j 0} \chi^k - j \delta_{k 0} \chi^j),
\end{align}
whereas the cobrackets are given by \eqref{d1},\eqref{d2}.
In particular in the Minkowski sector (defined by eqs.\eqref{pcduals} ) the non-vanishing commutators are
\begin{align}\label{minab}
[x^+, x^1]_A = \frac{n}{\kappa} x^+, \quad [x^-, x^1]_A = -\frac{n}{\kappa} x^- ,
\end{align}
i.e. in comparison with the lightcone $\kappa$-Poincar\'e the element $x^1$ plays the role of the advanced time coordinate $x^+$ and the scaling depends on the embedding. Note also that the Minkowski sector closes as an algebra.
\par
It is also possible to construct a quantum Lie-algebra by performing the deformation of the generators according to
\begin{align}
l_m^{\mathcal{F}_A} \equiv \mathcal{D}^{\mathcal{F}_A}(l_m) = l_m e^{\frac{im}{\kappa} T_0}, \\
T_m^{\mathcal{F}_A}  = T_m e^{\frac{im}{\kappa} T_0}.
\end{align}
It follows that the Casimir operator for any given embedding is undeformed
\begin{align}
\mathcal{C}_n^{\mathcal{F}_A} = 2 P_+^{\mathcal{F}_A} P_-^{\mathcal{F}_A} - \left(P_1^{\mathcal{F}_A} \right)^2 = \mathcal{C}_n
\end{align}
and thus all (scalar) particles have undeformed dispersion relation.
\newline
Using eq.\eqref{twcop} yields the twisted coproducts for the deformed generators
\begin{align}
\Delta_A (T_m^{\mathcal{F}_A}) =&  T_m^{\mathcal{F}_A} \otimes 1 + \exp \left( \frac{im}{\kappa} T_0^{\mathcal{F}_A} \right) \otimes T_m^{\mathcal{F}_A}, \\
\Delta_A(l_m^{\mathcal{F}_A}) =& l_m^{\mathcal{F}_A} \otimes 1 + \exp \left( \frac{im}{\kappa} T_0^{\mathcal{F}_A} \right) \otimes l_m^{\mathcal{F}_A} -\frac{im}{\kappa}  \exp \left( \frac{im}{\kappa} T_0^{\mathcal{F}_A} \right) l_0^{\mathcal{F}_A} \otimes T_m^{\mathcal{F}_A}.
\end{align}
Finally, the twisted brackets are given via eq.\eqref{fbrak} by
\begin{align}
[ l_m^{\mathcal{F}_A}, T_p^{\mathcal{F}_A}]_{\mathcal{F}_A} = & \;l_m^{\mathcal{F}_A} T_p^{\mathcal{F}_A} - T_p^{\mathcal{F}_A} l_m^{\mathcal{F}_A} + \frac{i}{\kappa} p \,m\, T_p ^{\mathcal{F}_A} T_m^{\mathcal{F}_A} \nonumber \\
= & \; (m-p) T_{p+m}^{\mathcal{F}_A}
\end{align}
and
\begin{align}
[l_m^{\mathcal{F}_A}, l_p^{\mathcal{F}_A}]_{\mathcal{F}_A} = & l_m^{\mathcal{F}_A} l_p^{\mathcal{F}_A} - l_p^{\mathcal{F}_A} l_m^{\mathcal{F}_A} + \frac{i}{\kappa} pm (T_p^{\mathcal{F}_A} l_m^{\mathcal{F}_A}- l_p^{\mathcal{F}_A} T_m^{\mathcal{F}_A}) - \frac{m^2 p^2}{\kappa} T_p^{\mathcal{F}_A} T_m^{\mathcal{F}_A} \nonumber \\
= & (m-p) l_{m+p}^{\mathcal{F}_A} - \frac{i}{\kappa} mp(m-p) T_{p+m}^{\mathcal{F}_A}.
\end{align}
Again we find that the structure constants are deformed similar to eqs.\eqref{res},\eqref{res2} with $n=0$ and $P^1  = i T_0$ playing the role of $P_+$.
\newline
Thus this twist seems physically reasonable and the deformed structures resemble those of the Jordanian part of the lightcone $\kappa$-Poincar\'e twist. Additionally the big advantage is that no embedding is singled out which seems to be more in line with the spirit of the BMS analysis. For an observer choosing coordinates $x^{\mu}$ there is still an explicit $n$ appearing in the $\kappa$-Minkowski relations \eqref{minab} which acts as a scaling factor for the characteristic length $1/\kappa$. This behavior may be  related to the rescaling of the ambient metric $\eta$ after eq.\eqref{pc4}.

\section{Bialgebra duals for BMS in 4d gravity}\label{sec6}

Having analyzed the three dimensional case we now return to the physically more relevant four dimensional case where we sketch the procedure of deforming and dualizing the BMS algebra.

We denote the basis of $\mathcal{B}^4$ as
\begin{align}
k_m = l_m + \bar l_m, \quad \bar k_m = -i (l_m - \bar l_m) , \quad S_{mp} = \frac{1}{2}(T_{mp} + T_{pm}), \quad A_{mp} = - \frac{i}{2}(T_{mp} - T_{pm})
\end{align}
and their respective duals satisfying
\begin{align}
\braket{k_m, \xi^n} & =  \delta_{m}^n, \quad \braket{\bar k_{\bar m}, \bar \xi^{\bar n}} = \delta_{\bar m}^{\bar n},\\
\braket{S_{pq}, \chi_S^{rs}} & = \frac{1}{4} \delta_{(p}^r\delta_{q)}^s, \quad \braket{A_{p'q'}, \chi_A^{r's'}} = \frac{1}{4} \delta_{[p'}^{r'}\delta_{q']}^{s'},
\end{align}
form a basis of $\mathcal{B}_4^*$ that is constructed similar to the three dimensional case as discussed in the appendix.
According to the embeddings discussed after eq.\eqref{genB} the family of inequivalent twists reads
\begin{align}
\mathcal{F}_{m,\epsilon}=& \exp \left(\frac{i\epsilon}{\sqrt{2}\kappa (1-2m)}( k_{1-2m} \otimes S_{1-m, m} - \bar k_{1-2m} \otimes A_{1-m, m} )\right) \nonumber \\
& \times \exp \left(- \frac{k_0}{(1-2m)} \otimes \log \left(1- \frac{i S_{1-m, 1-m}}{\sqrt{2}\kappa} \right) \right) \label{4dtwist}
\end{align}
which is parametrized by $m \in \mathbb{Z}$ and $\epsilon \in \{ 0, 1 \}$ denoting the Jordanian and full twist respectively. Thus we can in principle obtain the Hopf algebra $U\mathcal{B}^4_{m, \epsilon}$ with coproduct $\Delta_{m, \epsilon }$ and antipode $S_{m, \epsilon}$. The case $m=1$ has been considered more detailed in our previous publication \cite{Borowiec:2018rbr}.

Calculating the full deformed coproducts for all embeddings is tedious and for the construction of the Lie Bialgebra dual structures we only need the cobrackets which can be obtained from the $r$-matrix (cf. appendix \eqref{cobr41}-\eqref{cobr42})
Then, constructing the algebra structure in the ``Minkowski sector'' of  the dual of $\mathcal{B}^4_{r_{m,\epsilon}}$ analogous to the three dimensional case yields
\begin{align}
[x^1, x^+]_{m,\epsilon} &= \epsilon \frac{1}{\kappa} x^1, \quad [x^2, x^+]_{m,\epsilon} = \epsilon \frac{1}{\kappa} x^2, \quad [x^-, x^+ ]_{m,\epsilon} = \frac{1}{\kappa} x^-, \\
 [x^-, x^1]_{m,\epsilon} &= - \epsilon \frac{i\sqrt{2}}{\kappa} \chi_S^{3m-1, m}, \quad [x^-, x^2]_{m,\epsilon} = - \epsilon\frac{i\sqrt{2}}{\kappa} \chi_A^{3m-1, m}, \\
[x^1, x^2]_{m,\epsilon} &= -\epsilon \frac{i\sqrt{2}}{\kappa} \chi_A^{3m-1, 1-m} ,
\end{align}
 which also does only close on its own for the Jordanian ($\epsilon=0$) case.

We therefore find that from an (co)algebraic perspective the three and four dimensional duals of $\mathcal{B}_r$ have similar properties. Also, the calculations for the quantum Lie algebra carry over from the three dimensional setting if we consider the Jordanian and extended Jordanian twist.
Note, however, that the abelian twist can not be applied in four dimensions as there the only overlapping elements from the different embeddings are $l_0$ and $\bar l_0$ so the resulting twist would not correspond to a noncommutative spacetime of Lie algebra type that also does not admit a mass scale\footnote{The abelian twist using two rotations/boosts results in quadratic Minkowski relations, \cite{Lukier06}.}.

\section{Conclusion}\label{sec7}

The main goal of this paper is the analysis of deformations of the three and four dimensional BMS algebra as a quantum group. These deformations are initially defined in the Poincar\'e algebra and one finds that the consistent extension to the BMS requires deformations realized by a twist.
A crucial fact about the BMS algebra in three and four dimensions is that there are infinitely many subalgebras that are isomorphic to the Poincar\'e algebra, the so-called embeddings. We demonstrate in three dimensions that each of these subalgebras leaves invariant a different, distinct solution of the vacuum Einstein equations, which are shown to contain a conical singularity. Since the twist corresponding to the lightlike $\kappa$-Poincar\'e algebra as well as its Jordanian part are expressed in terms of Poincar\'e generators the existence of these subalgebras leads to infinitely many quantum groups with different coalgebra structures. They cannot be simultaneously realized which means that they break the symmetry between the embeddings and thus lead to a qualitatively different picture of the asymptotic symmetries.

In the duals of the Lie bialgebras that are the classical limit of these quantum groups we examined the Minkowski subalgebras which should, as in the Poincar\'e case, describe a non-commutative spacetime. For the extended Jordanian twist this subalgebra does not close and the physical interpretation of this fact is still not entirely clear. This comparison thus favors the Jordanian twist as the Minkowski subalgebra is identical to the dual of the deformed Poincar\'e algebra.

As all the embeddings overlap in certain elements one can construct another twist out of them which therefore does not single out an embedding and is abelian. Its Minkowski subalgebra also closes but it can be only constructed in three dimensions, whereas an abelian twist in four dimensions would not lead to a non-commuting spacetime of Lie algebra type which contains a mass scale $\kappa$ (cf. \cite{Lukier06}).

All of the analyzed types of twists are physically viable but they differ in potentially observable ways. Both the Jordanian and abelian twist admit an undeformed d'Alembert operator which corresponds to an unaltered dispersion relation. For the full lightcone $\kappa$-Poincar\'e this dispersion relation is deformed for massive and massless particles alike. Furthermore, all twists lead to different non-trivial coalgebra structures which dictate the addition of supermomenta.
This point is also relevant for the argument brought forward in \cite{Borowiec:2018rbr} ought to counter objections against the soft hair solution of the black hole information loss paradox. The non-trivial coproduct of the supermomenta could allow the total momentum to hold information of hard and soft modes combined non-linearly which would therefore not completely disentangle as the objection goes.

%\textcolor{red}{
	Last but not least it would be of interest to investigate a possible extension of the results presented here to the case of non-vanishing cosmological constant since quantum gravity is better understood in AdS spacetime. Asymptotic AdS spaces were first discussed in \cite{Henneaux:1985tv} and recently the $\Lambda$-extension of the BMS group/algebra was proposed in \cite{Compere:2019bua} (see also \cite{Fiorucci:2020xto}). Because the structure of the $\Lambda$-BMS algebra in four dimensions is in fact that of an algebroid it might not be straightforward to apply the tools discussed in this paper to deform it. Therefore that deformation provides a good direction for further research.  %.}

\section*{Acknowledgments}

This work is supported by funds provided by the Polish National Science Center (NCN), project UMO-2017/27/B/ST2/01902 and for LB, JKG, and JU also by the project number  2019/33/B/ST2/00050.

\appendix
\section{  $\cW(2,2)$ algebra} \label{AppendixA}

\par
$\cW$-type algebras form a general class of algebras extending the celebrated Witt algebra being  a simple infinite dimensional Lie algebra that can be identified with polynomial (holomorphic) vector fields on a circle. Both BMS algebras, in three and four spacetime dimensions, belong  to this class as well, see e.g. \cite{Sheikh} and references therein.

Here we are focused on the simplest case of real $\mathcal{B}^3$ algebra   with the Lie brackets
\begin{align}\label{ab1}
	[l_m, l_n ] = (m-n) l_{m+n}, \quad [l_m, T_n] = (m-n) T_{m+n}\,\quad m,n \in \mathbb{Z}.
\end{align}
In the mathematical literature it is a.k.a. $\cW(2,2)$-algebra \footnote{In some other convention it is also denoted as $\cW(0,-1)$ \cite{Sheikh}.} which is a semi-direct product of the centreless (two-sided Witt) algebra $\cW=span\{l_n; n \in \mathbb{Z}.\}$ with its module $\cT=span\{T_n; n \in \mathbb{Z}.\}$, i.e.
$ \cW (2,2)= \cW \ltimes  \cT$. Alternatively, it can be considered as a complex Lie algebra equipped with
an obvious anti-linear involutive anti-automorphism (hermitian conjugation):
$l_m^\dagger=-l_{m}, T_m^\dagger=-T_{m}$ \footnote{There is also an involutive (linear) automorphism $l_m\mapsto -l_{-m}, T_m\mapsto -T_{-m}$ of this algebra.}.
The last (complex algebra) point of view is most convenient for the purpose of quantum deformations.

There is quite extensive mathematical literature devoted to this subject. Therefore, several facts concerning $\cW(2,2)$ algebra are already established
(see e.g. \cite{Song:2019}, \cite{SongSu} - \cite{ChengSunYang}):
\begin{itemize}
	\item it is a perfect Lie algebra, i.e. $[\cW(2,2), \cW(2,2)]=\cW(2,2)$;
	\item it is centreless and admits universal central extension with two central generators;
	\item any maximal nonabelian finite-dimensional Lie subalgebra has dimension six and is isomorphic to $3D$ Poincar\'e algebra;
	\item its automorphism group is known;
	\item all Lie bialgebra structures are coboundary, triangular and of Jordanian type (not fully classified);
	\item some Jordanian-type Lie bialgebra structures and their duals are described;
	\item the corresponding Jordanian quantum deformations are described;
	\item its representations can be expressed in terms of vertex operator algebras;
	\item it admits a supersymmetric extension.
\end{itemize}

A remarkable property of the $\mathcal{B}^3$ algebra is that it contains (real) six-dimensional subalgebras $span\{l_k, T_k; k=-n,0,n\}$ which can be identify with the Poincar\'{e} algebra. They are maximal nonabelian finite-dimensional subalgebras.
On the other hand there are infinitely many infinite-dimensional non-isomorphic subalgebras. The most explored one is so-called half or one-sided $\cW(2,2)$ algebra defined usually as  $\cW(2,2)_1=span\{l_k, T_m; k, m \geq -1\}$.

In this paper we are interested in lightcone bialgebra structures induced on $\cW(2,2)$ by the embeddings \eqref{pc3} - \eqref{pc4}.
First, we notice that as it is well-known all Lie bialgebra structures of $\cW(2,2)$ are coboundary triangular and they are partially known \cite{SongSu,LiSuXin}. Therefore,  each such structure is implemented by a classical $r$-matrix satisfying the classical (unmodified) YB equation. In our case we are using $\kappa$ lightcone non-isomorphic $r$-matrices (cf. \eqref{cl_r_matrix}) \footnote{Notably, their elements belongs to $\cW(2,2)_1$ subalgebra.}
\begin{equation}\label{ab4}
	r_{n,\epsilon}= \frac{1}{\sqrt{2}n} \left( l_0 \wedge T_n + \epsilon\, l_n \wedge T_0 \right)
\end{equation}
belonging to a class of extended Jordanian twist (see e.g. \cite{BP2} and references therein),
where $\epsilon=0,1$. Jordanian case $\epsilon=0$ has been studied in the literature together corresponding quantum deformations of the enveloping algebra of $\cW(2,2)$  \cite{SongSuWu,LiSu}. To our knowledge, extended deformations of $\cW(2,2)$ have not been studied yet.
%\begin{align}\label{cl_r_matrix}
%	\delta_{n, r} (X) = ad_{r_n} \triangleright X, \quad r_n =g^{- +} M_{+ -} \wedge P_+ + \epsilon\, g^{1 1} M_{+1} \wedge P_1 = \frac{1}{\sqrt{2}n} \left( l_0 \wedge T_n + \epsilon\, l_n \wedge T_0 \right)
%\end{align}
We recall that although these bialgebra structures and the corresponding deformations are equivalent on the level of Poincar\'{e} subalgebras there are not equivalent on the level of infinite-dimensional  $\cW(2,2)$  algebra. This is so, since they cannot be transformed each other by a Lie algebra automorphism
of $\cW(2,2)$, which in the most general form can be described by the formula:
\begin{equation}\label{ab4}
	l_n\mapsto a^n(\omega l_{\omega n} + n b T_{\omega n})\,,\quad T_n\mapsto a^n c T_{\omega n}\,,
\end{equation}
where $a, b, c\in \mathbb{C}$, $a,c \neq 0$ and $\omega
=\pm 1$. In particular, $T_n$ cannot be transformed to $T_m$ with $m\neq\pm n$.

Besides \cite{LiSuXin}, there are also other possible candidates for triangular deformations
of $\cW(2,2)$ algebra by twists associated with Poincar\'e subalgebras. The classification of the corresponding $r$-matrices can be found in \cite{Lukier}.

 \section{Bialgebra duals}\label{AppendixB}

 When constructing the Lie-bialgebra that is dual to the Lie bialgebras $\mathcal{B}^3_{r_{n,\epsilon}}, n=1,2,\ldots\in\mathbb{N}, \epsilon\in\{0,1\}$ one faces the problem that in an infinite-dimensional vector space $V$ we do not necessarily have $(V \otimes V)^* = V^* \otimes V^*$ and thus the cobracket $\delta^{\circ} \equiv [\; ,\;]^* : V^* \rightarrow (V \otimes V)^*$ might not be closed. In order to fix this, one can define a subspace $V^{\circ} \subseteq V^*$ as a sum of all good subspaces, i.e. subspaces $\mathcal{S}$ of $V^*$ such that $[\;, \;]^* (S) \subset \mathcal{S} \otimes \mathcal{S}$. In our case we can make use of the fact that the generators of the BMS can be expressed in terms of monomials in an abstract commuting variable $x$ (and its formal inverse), i.e. as a Laurent polynomials algebra $\mathbb{C}[x^{\pm1}]$,
 \begin{align}\label{b1}
 	l_m \equiv (x^{m+1}, 0), \quad T_m \equiv (0, x^{m+1}).
 \end{align}
 Since $\mathbb{C}[x^{\pm1}]$ is associative and commutative one can apply the following theorem (for details see \cite{Taft1,Taft2,Griffing}).
 \newline
 \textbf{Theorem:} Consider $A\equiv\mathbb{C}[x^{\pm1}]$ as a complex  associative and commutative algebra with the linear space basis consisting of monomials $\{x^j|j \in \mathbb{Z}\}$. Its algebraic dual  $A^*$ admits pseudo-basis $\{\varepsilon^j|j \in \mathbb{Z}\}$
 \footnote{One should notice that any element $a\in A$ can be decomposed as $a=\sum_{n\in \mathbb{Z}} a_n x^n$ with only finite number non-vanishing components $(a_n)_{n\in \mathbb{Z}}$. Instead, for $f\in A^*$ one writes
 	$f=\sum_{n\in \mathbb{Z}} f_n \varepsilon^n$ without any restrictions.
 }. Then $A^{\circ}$ is the subspace of all sequences $(f_n)_{n\in \mathbb{Z}}$ satisfying linearly recursive relations in the form $f_n = \sum_{j=1}^r c_j f_{n-j}$ for some  $1\leq r \in \mathbb{N} $ and some coefficients $c_1\ldots c_r \in \mathbb{C}$, $c_r\neq 0$. In this  case the elements $f_0,f_1, \ldots, f_{r-1}\in\mathbb{C}$ serve as initial conditions. %for $n>0$ and  $f_0,f_{-1}, \ldots, f_{1-r}$ for $n<0$.
 The same is true for $A$ treated as  the Witt algebra with the Lie bracket $[l_n, l_m]=(m-n)l_{n+m}; n, m\in \mathbb{Z}$, where the generator $l_n$ can be identify with the monomial $x^{n+1}$ throughout first order differential operator realisation $l_n=x^{n+1}d/dx$.\smallskip\\
 %\newline

 The case of the Witt algebra is most interesting for us.
 %$A=\cW_1: [x^n, x^m]=(m-n)x^{n+m}; n, m\geq -1$.
 It has been found in \cite{Taft1,Taft2} that the algebra $A^\circ$ has a basis \footnote{ In fact, the elements $\varepsilon^n$ do not belong to $A^\circ$. However, they are linearly independent in the half Witt algebra dual $\cW_1^\circ$ spanning finite sequences, where $\cW_1$ is defined by the relations $[l_n, l_m]=(m-n)l_{n+m}; n, m\geq -1$. In such a case the elements $\varepsilon^n\,,\,n\geq -1$ together with $\varepsilon^{f,j}=(f^n n^j)_{n\geq -1}$ form a basis in $\cW_1^\circ$.}
 \begin{align}\label{b2}
 	\varepsilon^{f,j}=(f^n n^j)_{n\in \mathbb{Z}}
 \end{align}
 where $0\neq f\in \mathbb{C}$ and $j=0,1,\dots\in \mathbb{N}$. For example, for $j=0$ one gets the geometric series $\varepsilon^{f,0}=(f^n)_{n\in \mathbb{Z}} $. The duality relations reads now as
 \begin{align}\label{b3}
 	<\varepsilon^{f,j}, x^n>= f^n n^j \in\mathbb{C}\,,\quad \mbox{for each}\ \   n\in \mathbb{Z}
 \end{align}
 In the same papers, a cobracket  $\delta^\circ :A^\circ\rightarrow A^\circ\otimes A^\circ$ which is dual to the Witt Lie bracket in $A$ have been found in the form
 \begin{align}\label{b4}
 	\delta^\circ(\varepsilon^{f,i})=\sum_{j=0}^{j=i+1}\left[\binom{i}{j}-\binom{i}{j-1}\right] \varepsilon^{f,j}\otimes\varepsilon^{f,i+1-j}\,.
 \end{align}
 Particularly, for a geometric series $\delta^\circ(\varepsilon^{f,0})=\varepsilon^{f,0}\otimes\varepsilon^{f,1}-\varepsilon^{f,1}\otimes\varepsilon^{f,0}$. For the one-sided Witt algebra dual $\cW_1^\circ$ in addition to the relations \eqref{b4} one gets
 \begin{align}\label{b5}
 	\delta^\circ(\varepsilon^m)= \sum_{i+j = m} (j-i) \varepsilon^i \otimes \varepsilon^j = \sum_{i = -1}^{m+1} (m-2i) \varepsilon^i \otimes \varepsilon^{m-i}\,,
 \end{align}
which for $i,j,m\geq -1$ yields the finite formula.

 Concerning bialgebra duals also the case of $\cW_1$ is better understood. All bialgebra structures are triangular  and fully classified by means of the following Jordanian $r$-matrices: \footnote{All bialgebra structures on two-sided Witt algebra are also triangular of Jordanian type (including \eqref{b6}) but not fully classified.}
 \begin{equation}\label{b6}
 	r_n=  l_0 \wedge l_n \,.\,
 \end{equation}
that imply coalgebra brackets in $\cW_1$
 \begin{equation}\label{b7}
 \delta_n(l_m) = (m-n) l_0 \wedge l_{n+m} + m\, l_m \wedge l_n\,.\end{equation}
 The dual family of Lie algebra brackets in  $\cW_1^\circ$ reads now \cite{Taft2,Zhifeng}
 \begin{align}
 	 [\varepsilon^0,\varepsilon^m]_n= &  ( m-2n)\varepsilon^{m-i}\,,\quad \mbox{for}\quad m\neq 0\\
   [\varepsilon^m,\varepsilon^n]_n= &  m\,\varepsilon^{m}\,,\quad\quad \mbox{for}\quad\quad m\neq 0, n\,.
 \end{align}
 %\newline

 Thus, in Sec. 4, since $\cW(2,2)\sim A \oplus A$ then $\cW(2,2)^\circ\sim A^\circ \oplus A^\circ$ as a linear spaces.
 For $A \equiv \mathbb{C}[x^{\pm1}]$ and $\mathcal{B}^3 \equiv  A \oplus A$ one has $\mathcal{B}^{3\,\circ} = A^{\circ} \oplus A^{\circ}$ and we denote the pseudo-basis of $\mathcal{B}^{3*}$  as $\{(\zeta^j, \chi^k), j, k \in \mathbb{Z} \}$.
 The bracket and cobracket structure of the Lie Bialgebra $\mathcal B^{3\, \circ}_{r_n}$, is defined to be dual to $\mathcal B^3_{r_n}$.

 \section{Cobrackets for Lie bialgebra structures on BMS 4}\label{AppendixC}

The classical $r$-matrices corresponding to a family of twists \eqref{4dtwist} read
\begin{align}
r_{n, \epsilon}= & \eta^{+-} M_{+-} \wedge P_+ + \epsilon \eta^{aa} M_{+a} \wedge P_a \\
= & \frac{1}{\sqrt{2}(1-2m)}\left(- k_0 \wedge S_{1-m, 1-m} + \epsilon (k_{1-2m} \wedge S_{1-m, m} +  \bar k_{1-2m} \wedge A_{1-m, m} ) \right).
\end{align}

Thus the corresponding bialgebra cobrackets take the form
\begin{align}
\delta_{n, \epsilon}(k_m) = \frac{i m}{\sqrt{2} (1-2n)} k_m \wedge S_{1-n, 1-n} -\frac{2 i }{\sqrt{2} (1-2n)} k_0 \wedge \left( \frac{m+1}{2} -(1-n) \right) S_{1-n + m , 1-n} \nonumber \\
+ \frac{i \epsilon }{\sqrt{2}  (1-2n)} (1-2n -m) ( k_{1-2n +m} \wedge S_{1-n, n} + \bar k_{1-2n +m} \wedge A_{1-n, n})  \nonumber \\
- \frac{i \epsilon }{\sqrt{2}  (1-2n)} k_{1-2n}  \wedge  \left( \left( \frac{m+1}{2} - (1-n) \right) S_{1-n+m, n} + \left( \frac{m+1}{2} -n \right) S_{1-n, m+n} \right) \nonumber \\
+ \frac{i \epsilon }{\sqrt{2}  (1-2n)} \bar k_{1-2n}  \wedge  \left( \left( \frac{m+1}{2} - (1-n) \right) A_{1-n+m, n} + \left( \frac{m+1}{2} -n \right) A_{1-n, m+n} \right), \label{cobr41}
\end{align}
\begin{align}
\delta_{n, \epsilon}(\bar k_m) = \frac{i m}{\sqrt{2}\kappa (1-2n)} \bar k_m \wedge S_{1-n, 1-n} -\frac{2 i }{\sqrt{2} (1-2n)} k_0 \wedge \left( \frac{m+1}{2} -(1-n) \right) A_{1-n + m , 1-n} \nonumber \\
+ \frac{i \epsilon }{\sqrt{2}  (1-2n)} (1-2n -m) (\bar k_{1-2n +m} \wedge S_{1-n, n} - k_{1-2n +m} \wedge A_{1-n, n})  \nonumber \\
- \frac{i \epsilon }{\sqrt{2} (1-2n)} k_{1-2n}  \wedge  \left( \left( \frac{m+1}{2} - (1-n) \right) A_{1-n+m, n} + \left( \frac{m+1}{2} -n \right) A_{1-n, m+n} \right) \nonumber \\
+ \frac{i \epsilon }{\sqrt{2}  (1-2n)} \bar k_{1-2n}  \wedge  \left(- \left( \frac{m+1}{2} - (1-n) \right) S_{1-n+m, n} + \left( \frac{m+1}{2} -n \right) S_{1-n, m+n} \right),
\end{align}

\begin{align}
\delta_{n, \epsilon}(S_{pq}) = - \frac{i}{\sqrt{2} (1-2n)} (1-(p+q)) S_{pq} \wedge S_{1-n, 1-n} \nonumber \\
+\frac{i \epsilon}{\sqrt{2} (1-2n)} \left( (1-n -p) S_{p+1-2n, q} + (1-n-q) S_{q + 1-2n, p} \right) \wedge S_{1-n, n} \nonumber \\
-\frac{i \epsilon}{\sqrt{2}(1-2n)} \left( (1-n -p) A_{p+1-2n, q} + (1-n-q) A_{q + 1-2n, p} \right) \wedge A_{1-n, n},
\end{align}

\begin{align}
\delta_{n, \epsilon}(A_{pq}) = - \frac{i}{\sqrt{2} (1-2n)} (1-(p+q)) A_{pq} \wedge S_{1-n, 1-n} \nonumber \\
+\frac{i \epsilon}{\sqrt{2} (1-2n)} \left( (1-n -p) A_{p+1-2n, q} + (1-n-q) A_{q + 1-2n, p} \right) \wedge S_{1-n, n} \nonumber \\
-\frac{i\epsilon}{\sqrt{2} (1-2n)} \left( -(1-n -p) S_{p+1-2n, q} + (1-n-q) S_{q + 1-2n, p} \right) \wedge A_{1-n, n}. \label{cobr42}
\end{align}

\end{document}